\documentclass[article]{aa} 

\usepackage{graphicx}
\usepackage[none]{hyphenat}
\usepackage{txfonts}
\usepackage{gensymb}
\newcommand{\Ha}{$\mathrm{H}\alpha$\xspace}
\newcommand{\Hb}{$\mathrm{H}\beta$\xspace}
\newcommand{\NII}{$[\mathrm{N}\textsc{ii}]$\xspace}

\newcommand{\NIIb}{$[\mathrm{N}\textsc{ii}]\,\lambda 6584$\xspace}
\newcommand{\OIII}{$[\mathrm{O}\textsc{iii}]$\xspace}
\newcommand{\SII}{$[\mathrm{S}\textsc{ii}]$\xspace}
\newcommand{\OIIIb}{$[\mathrm{O}\textsc{iii}]\,\lambda 5007$\xspace}
\newcommand{\HeII}{$\mathrm{He}\textsc{II}\,\lambda 4686$\xspace}

\defcitealias{rodriguez_del_pino_properties_2019}{RDP19}

\begin{document} 

   \title{Extreme gas kinematics in an off-nuclear $\mathrm{H II}$\xspace region of \\SDSS J143245.98+404300.3}


   \author{Bruno~Rodr\'iguez~Del~Pino\inst{1},         
          Santiago~Arribas\inst{1},  
          Javier~Piqueras L\'opez\inst{1},\\
          Alejandro~Crespo G\'omez\inst{1} and
          Jos\'e~M.~V\'ilchez \inst{2} 
}
   \authorrunning{Rodr\'iguez Del Pino, et al. }

   \institute{Centro de Astrobiolog\'ia (CSIC-INTA), Torrej\'on de Ardoz, Madrid, Spain\\
             \email:{brodriguez@cab.inta-csic.es}
   \and{Instituto de Astrof\'isica de Andaluc\'ia. (CSIC). Apdo. 3004, 18080, Granada, Spain}
           }

   \date{Accepted for publication in Astronomy \& Astrophysics}

  \abstract{We present and discuss the properties of an ionized component with extreme kinematics in an off-nuclear $\mathrm{H II}$\xspace region located at $\sim0.8-1.0$~kpc from the nucleus of SDSS~J143245.98+404300.3, recently reported in \citet{rodriguez_del_pino_properties_2019}. The high-velocity gas component is identified by the detection of very broad emission wings in the \Ha line, with Full Width at Half Maximum ($FWHM$)~$\geq$~$850-1000$~kms$^{-1}$. Such gas kinematics are outstandingly high compared to other $\mathrm{H II}$\xspace~regions in local galaxies and similar to those reported in some star-forming clumps of galaxies at $z\sim2$. The spatially resolved analysis indicates that the high velocity gas extends at least $\sim90$~pc and it could be compatible with an ionized outflow entraining gas at a rate $\sim7-9$ times larger than the rate at which gas is being converted into stars. We do not detect broad emission wings in other emission lines such as \Hb, maybe due to moderate dust extinction, nor in $[\mathrm{N}\textsc{ii}]$$\lambda\lambda$6548,~6584 and $[\mathrm{S}\textsc{ii}]$$\lambda\lambda$6717,~6731, which could be caused by the presence of turbulent mixing layers originated by the impact of fast-flowing winds. The lack of spectral signatures associated to the presence of Wolf-Rayet stars points towards stellar winds from a large number of massive stars and/or supernovae as the likely mechanisms driving the high velocity gas. }
  
   \keywords{star formation --
                outflows --
                kinematics and dynamics 
               }

   \maketitle
%

\section{Introduction}

The presence of ionized gas moving at very high velocities in galaxies (Full Width at Half Maximum, $FWHM$~$\geq$~$1000$~kms$^{-1}$) is generally associated to active galactic nuclei (AGN)  and nuclear starbursts \citep[e.g.,][]{villar_martin_triggering_2014, heckman_implications_2016}, although they have also been identified in \rm{H~II}~galaxies \citep[e.g.,][and references therein]{terlevich_high-velocity_2014}, specially in Green Peas \citep{amorin_complex_2012}. However, gas moving at such high velocities has also been observed in off-nuclear \rm{H~II}~regions, although only a handful of cases have been reported so far in nearby galaxies \citep[e.g.,][]{diaz_detailed_1987, castaneda_remarkable_1990, roy_superbubble_1991}. Several mechanisms have been suggested to be responsible for the extreme kinematics observed in these off-nuclear regions, generally invoking the presence of a large number of massive stars and/or supernovae; however, there is still no consensus on their origin \citep{roy_origin_1992, binette_broad_2009}. Given the low number of detections (either by their low incidence or the observational limitations), the discovery of new ones and their detailed study is crucial to understand the physical mechanisms associated to them. Moreover, the fact that ionized outflows appear to be common in off-nuclear star-forming regions of galaxies at $z\sim2$ \citep{genzel_sins_2011} make their detections in nearby galaxies an important baseline for their study at high redshift.

The standard method to identify gas components with extreme kinematics in the spectra of galaxies is the detection of broad emission wings in lines such as \Ha, \Hb, \OIIIb or those associated to Wolf-Rayet stars \citep{diaz_detailed_1987, izotov_spectrophotometry_1996}. However, observing them is challenging because they normally appear as low-luminosity pedestals superimposed on the emission from the host galaxy. In addition, since most spectroscopic observations generally target the central parts of galaxies, regions located away from them are systematically missed. In this sense, the advent of large Integral Field Spectroscopic (IFS) surveys provide the appropriate spatial and spectral coverage to extend the search for these high velocity gas components to a large number of galaxies.

In our recent work \citet[][hereafter RDP19]{rodriguez_del_pino_properties_2019}, where we explored the properties of ionized outflows in a parent sample of $>2700$ galaxies from the Second Data Release \citep[DR2; ][]{abolfathi_fourteenth_2018} of the Mapping Nearby Galaxies at Apache Point Observatory survey \citep[MaNGA;][]{bundy_overview_2015}, we discovered extreme high velocity gas motions ($FWHM$~$\geq$~$1000$~kms$^{-1}$) in an off-nuclear \rm{H~II}~region located at $\sim$~1~kpc from the center of the galaxy SDSS~J143245.98+404300.3, at $z=0.0174$. The discovery was a result of the analysis of more than 5000 \Ha-emitting regions in $\sim1200$ systems, detecting signatures of ionized outflows in 105 individual regions. Here we analyze in more depth the MaNGA data and present new observations of this region performed with the Intermediate Dispersion Spectrograph and Imaging System (ISIS) on the 4.2m William Herschel Telescope (WHT) of the Isaac Newton Group (ING) at the Roque de los Muchachos Observatory (La Palma, Spain). Throughout this work we adopt a  $H_{\rm0}$~=~67.3~kms$^{-1}$~Mpc$^{-1}$, $\Omega_{\rm M}$~=~0.315, $\Omega_{\Lambda}$~=~0.685 cosmology \citep{planck_collaboration_planck_2014}.

\begin{figure}
	\begin{center}
		\includegraphics[width=0.49\textwidth]{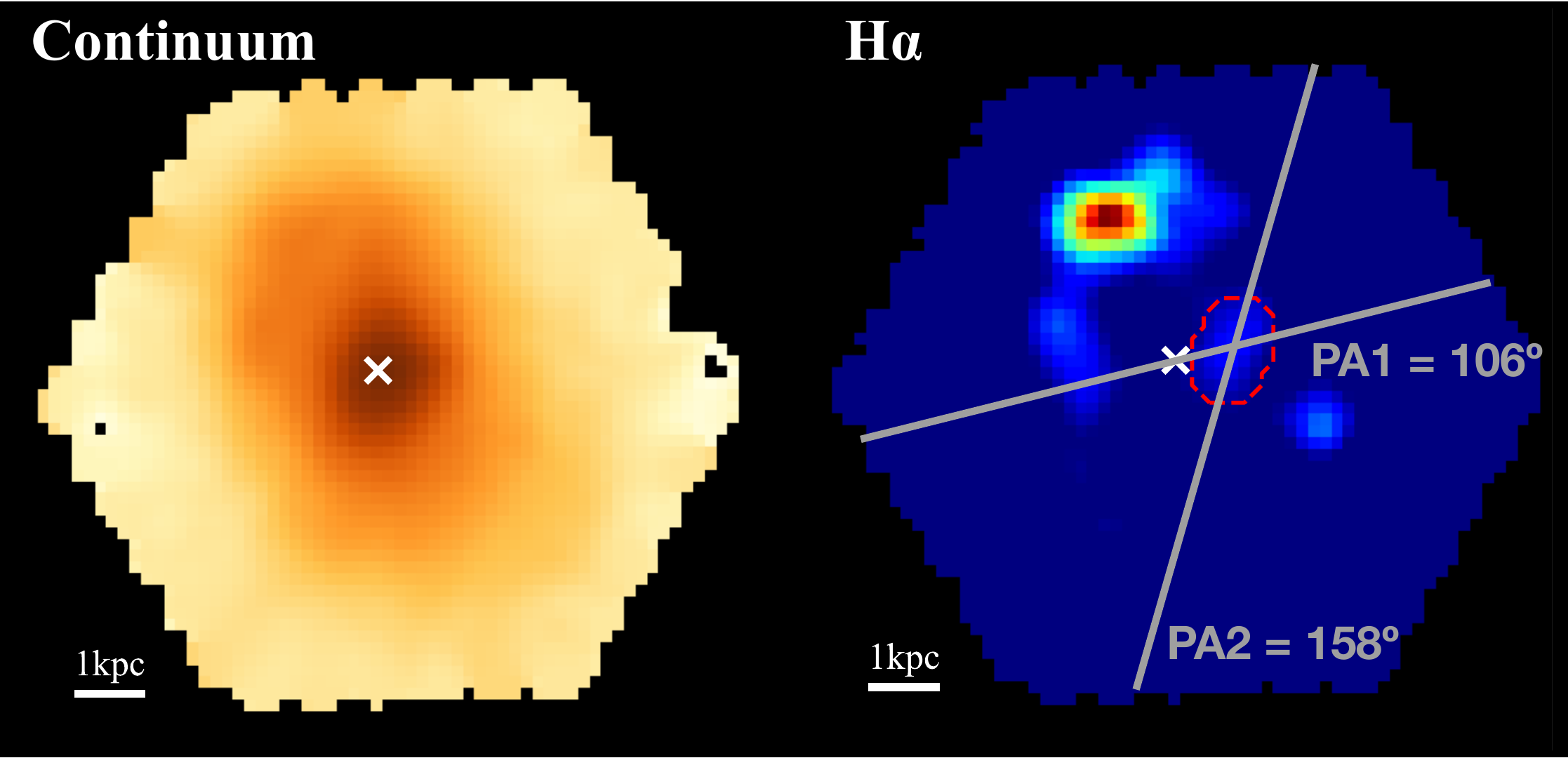}
		\caption[Halpha maps]{Continuum (\emph{left}) and \Ha (\emph{right}) emission maps of SDSS~J143245.98+404300.3 obtained from the MaNGA data. The red dashed line highlights the position of the off-nuclear \rm{H~II}~region where the high velocity gas component has been found. The white cross indicates the photometric center of the galaxy. We have also highlighted in grey the two positions used for the long-slit spectroscopic observations with ISIS/WHT.  
}
		\label{Halphamap}
	\end{center}
\vspace*{-20pt}	
\end{figure}

\section{Data and analysis}
\label{data}

In this work we use the MaNGA DR2 data of the galaxy SDSS~J143245.98+404300.3, with a wavelength coverage from 3600\AA{} to 10000\AA{} and a resolving power ranging from $R\sim1400$ to $R$~$\sim2600$ at short and long wavelengths, respectively. The spatial resolution element in the datacube is 2.5~arcsec~(FWHM), which corresponds to $0.875$~kpc at the redshift of the galaxy ($z$~=~0.0174). In Figure~\ref{Halphamap} we show the \Ha emission and nearby continuum maps of the galaxy generated as in \citetalias{rodriguez_del_pino_properties_2019}, highlighting (red, dashed line) the location of the off-nuclear \rm{H~II}~region. 

We also present and analyze follow-up long-slit spectroscopic observations taken with ISIS/WHT during the nights of March 9th and 10th of 2019. We used simultaneously the blue and red arms of ISIS with a 1.3 arcsec slit and the R600 grating, covering the wavelength ranges [4328-5584]~\AA{} and [6392-7594]~\AA{}. The spectral resolution ($FWHM$) measured on the sky lines are 0.95~\AA{} for the blue and 0.91~\AA{} for the red arm. The spectra were taken in several exposures of 40 mins each in two position angles (Figure~\ref{Halphamap}). The reduction of the ISIS/WHT data was performed using standard \textsc{IRAF} tasks. For the analysis performed in this paper we use the five exposures\footnote{In the rest of exposures either the seeing conditions were not optimal or the slit was not placed correctly on the position of the clump.} where the broad component in the \Ha line shows a more clear detection (three with PA1 and two with PA2). When these observations are combined, the total exposure time is $\sim3.33$ hours. We do not perform absolute flux calibration in the ISIS/WHT data because no standard star observation was included in the programme.

\begin{figure}
	\begin{center}
		\includegraphics[width=0.48\textwidth]{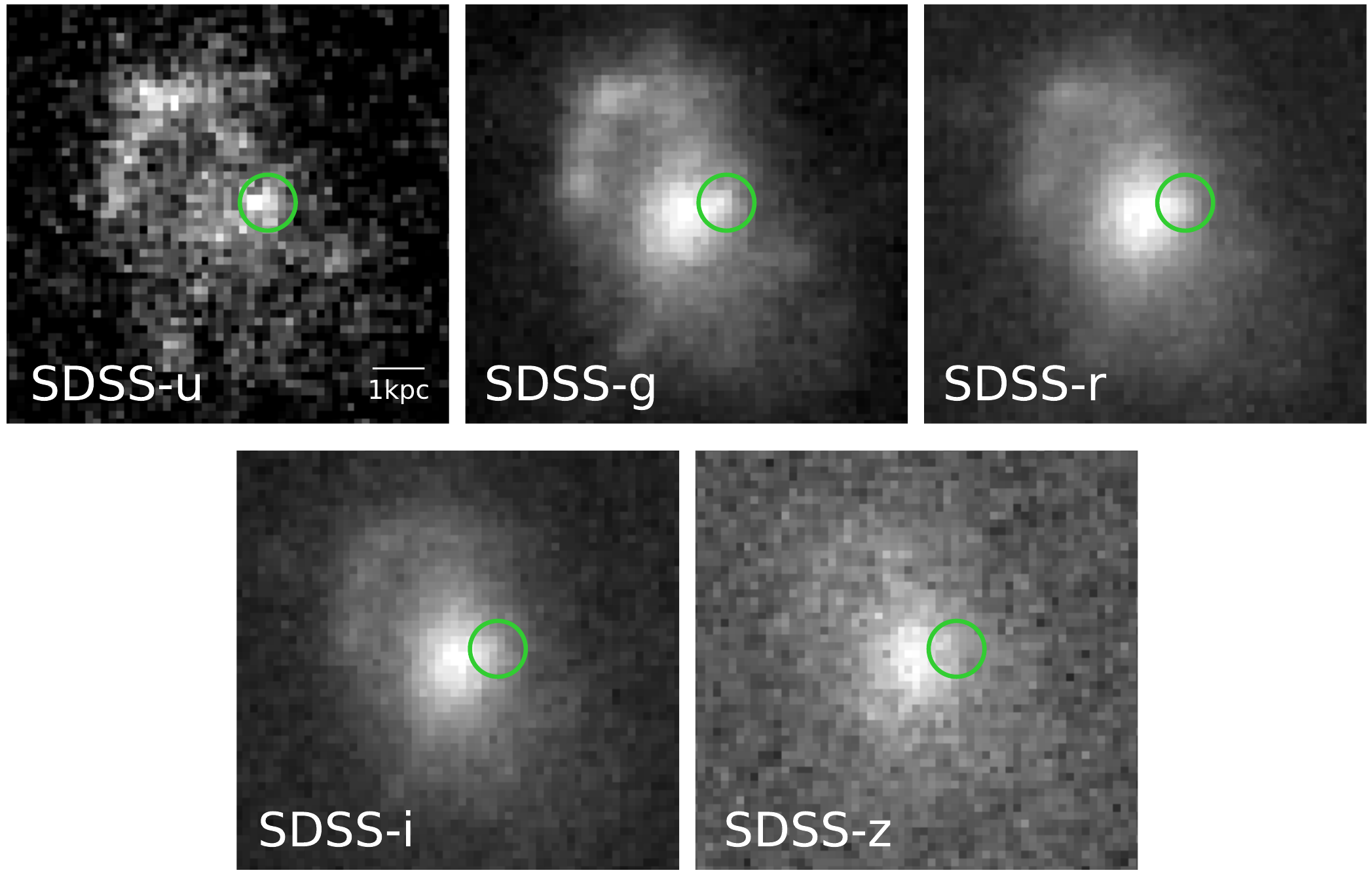}
	\caption[Spectral fit]{SDSS \emph{ugriz} images of SDSS J143245.98+404300.3 where we have highlighted the off-nuclear \rm{H~II}~region hosting the ionized gas at high velocities.}
	\label{SDSSimages}
	\end{center}
\end{figure}

\begin{figure}[b]
	\begin{center} 
		\includegraphics[width=0.49\textwidth]{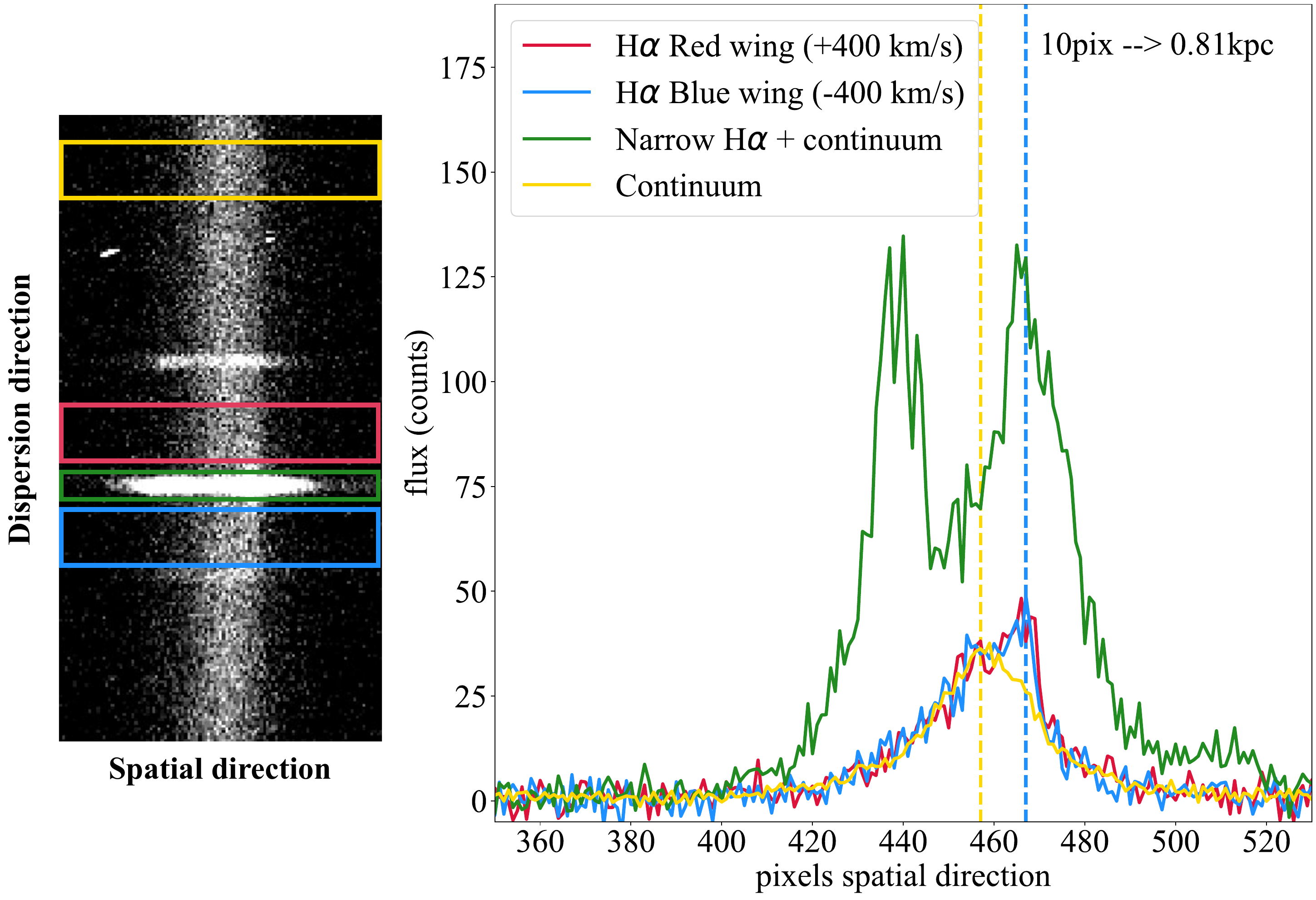}
		\caption[WHT distance to nucleus]{\emph{Left}: An example of a 2D reduced image from the ISIS/WHT data obtained with the slit in the PA1 orientation. The coloured regions are chosen to contain the light from the narrow component of \Ha (green), the continuum emission (yellow) and gas at velocities of $-400$~kms$^{-1}$ (blue) and $+400$~kms$^{-1}$ (red) with respect to the narrow component of \Ha (systemic). \emph{Right}: median light distribution for each of the regions selected in the \emph{left} image. The vertical, dashed lines mark the positions of the peaks of emission in the continuum and at $\pm400$~kms$^{-1}$ with respect to the systemic emission.}
		\label{wht_distance_nucleus}
	\end{center}
\vspace*{-20pt}	
\end{figure}

\begin{figure*}
	\begin{minipage}[h]{0.95\textwidth}
		\includegraphics[width=0.49\textwidth]{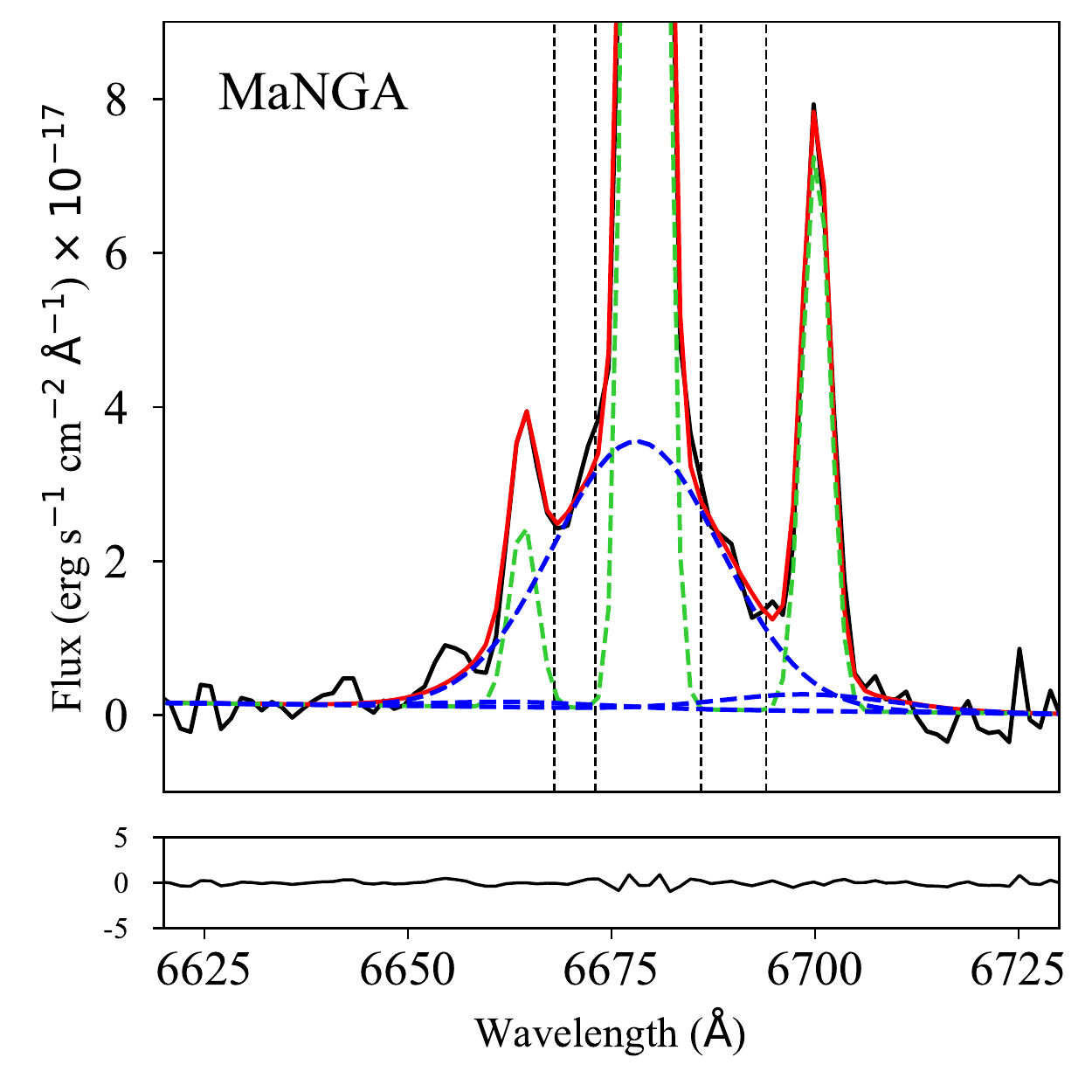}
		\includegraphics[width=0.49\textwidth]{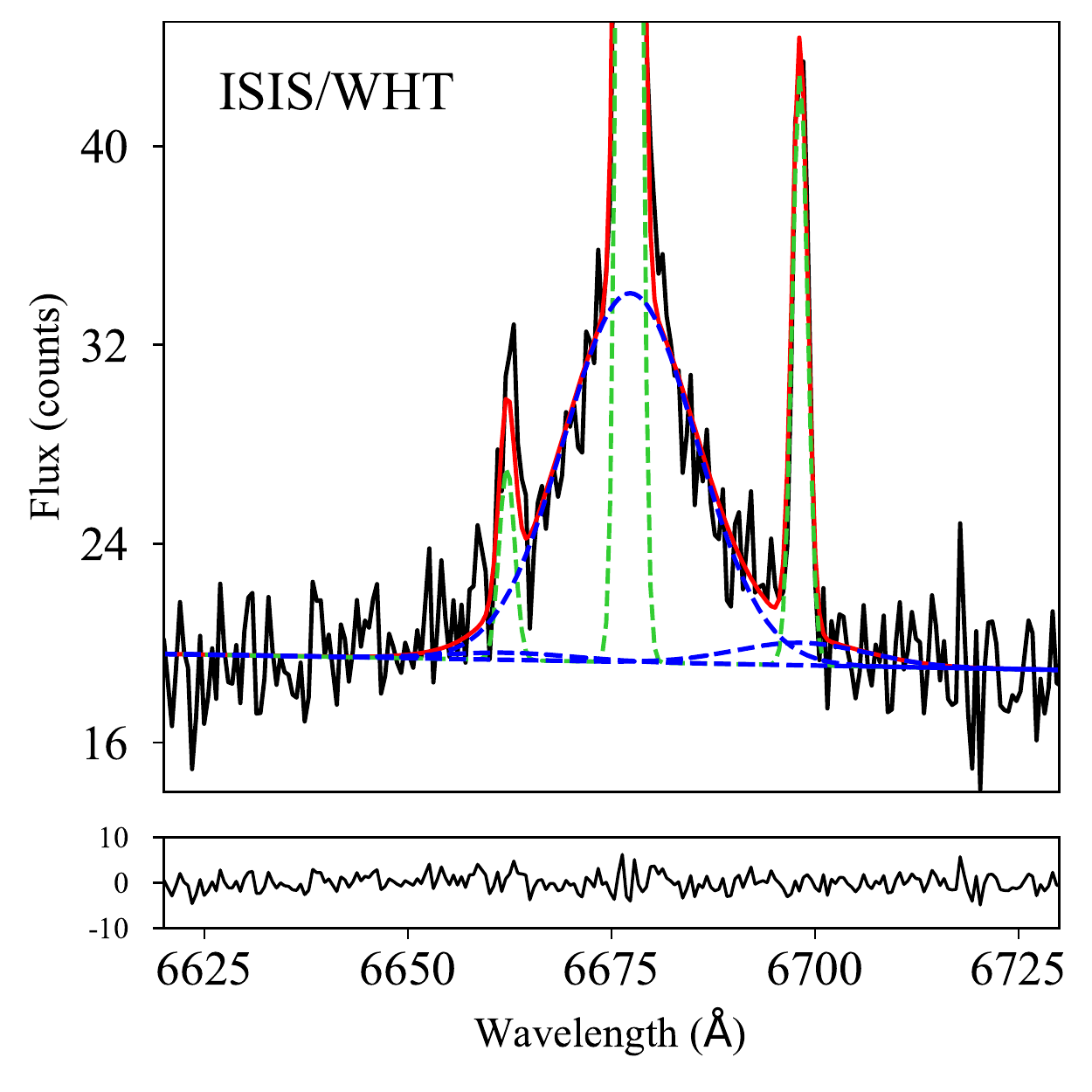}
	\end{minipage}
	\caption[Spectral fit]{Modelling of the MaNGA and ISIS/WHT spectra from the off-nuclear \rm{H~II}~region identified in SDSS~J143245.98+404300.3. Black, solid lines represent the observed data. Green and blue dashed lines correspond to the narrow and broad components, respectively, and the red, solid line is the sum of these two. The dashed, vertical lines mark the wavelengths ranges that are used to study the spatial extent of the broad emission in the MaNGA data (see Section~\ref{subsection_spatial}).}
	\label{spectral_fit}
\end{figure*}

\subsection{Isolating the region hosting the broad emission}
\label{subsec_isolating}

The identification of the off-nuclear \rm{H~II}~region in the MaNGA data was carried out in \citetalias{rodriguez_del_pino_properties_2019} by searching for individual \Ha-emitting regions in the host galaxy. The center of the region is located at $\sim1$~kpc from the photometric center of the galaxy, measured by fitting a 2D Gaussian to the continuum emission in a region bluewards \Ha (see Figure~\ref{Halphamap}). We extracted and combined the spectra from the \Ha-emitting region hosting the high velocity gas using an aperture radius of $\sim$2~arcsec (0.74~kpc), corresponding to an area of $\sim$~13~arcsec$^2$. In Figure~\ref{SDSSimages} we show the SDSS \emph{ugriz} images of SDSS~J143245.98+404300.3, highlighting the location of the off-nuclear \rm{H~II}~region identified in the MaNGA data. The images in the \emph{u} and \emph{g} bands (perhaps also in the \emph{r} band) show evidence for significant emission associated to the \rm{H~II}~region.

To isolate the broad emission in the ISIS/WHT long-slit spectroscopic data we first extract the light profiles at different positions in the dispersion direction from the reduced, 2D images, as shown in the \emph{left} panel of Figure~\ref{wht_distance_nucleus}. In this figure we represent the case of a single exposure taken with the slit in the PA1 orientation, where the center of the galaxy is also included (see Figure~\ref{Halphamap}). We have selected different regions in the dispersion direction, whose median light profiles are shown in the \emph{right} panel. In this plot a clear peak of emission can be seen at high velocities (red and blue) that is shifted $\sim10$~pixels ($\sim$~2.2~arcsec) with respect to the peak of continuum emission (yellow), corresponding to the central parts of the galaxy. Such spatial offset in pixels corresponds to $0.81$~kpc, similar to the distance measured in the MaNGA data ($\sim1$~kpc) and confirming the off-nuclear location of the broad emission. In this figure one can also see the emission from the other \Ha-emitting region in the opposite side of the galaxy that is also included in the slit. For the five exposures with better detections of the broad component, we extract and combine the spectra from the spatial positions (4-5 pixels) associated to the broad emission, which taking into account the width of the slit and the spatial scale (0.22~arcsec/pixel), correspond to extraction areas of $\sim1-1.4$~arcsec$^2$. 

\subsection{Kinematics of the ionized gas}

We focus now on the study of the emission line profiles to constrain the kinematics of the ionized gas. For this analysis the spectra need to be corrected for stellar absorption, which can be specially severe in the \Hb line. For the MaNGA data, the modelling and subtraction of the stellar contribution was already done in \citetalias{rodriguez_del_pino_properties_2019}. However, in the case of the non flux calibrated ISIS/WHT data, the continuum fitting is very uncertain due to the different flux levels of the blue and red arms of the ISIS instrument. Therefore, for the analysis of the blue part of the spectra, which include the \Hb and \OIIIb lines, we will only use the MaNGA data, whereas for the red part, including \Ha, \NII and \SII, we use both datasets. For the analysis of the \Ha line we expect the effects from stellar absorption to be negligible.

We start by fitting the \Ha and the neighbouring $[\mathrm{N}\textsc{ii}]$$\lambda\lambda$6548,~6584 emission lines with a model consistent of two kinematic components: a narrow one to reproduce the systemic rotation of the host galaxy and a broad one to constrain the properties of the gas moving at high velocities. As in \citetalias{rodriguez_del_pino_properties_2019}, we perform the modelling of the spectra using a Bayesian approach based on Markov Chain Monte Carlo (MCMC) techniques, developed in the software \textsc{emcee} \citep{foreman-mackey_emcee:_2013}. The value adopted for each model parameter corresponds to the median of its probability density distribution and the errors are taken from the 16th and 84th percentiles. 

In Figure~\ref{spectral_fit} we show the results from the spectral modelling\footnote{Note that due to the larger extraction apertures employed (see Section~\ref{subsec_isolating}), the S/N in the MaNGA data is higher and the narrow lines are broader (a larger area encompasses higher dispersion velocities).}. In the MaNGA data, $FWHM_{\rm broad}$ is $1074$$\pm$$160$~kms$^{-1}$, whereas in the ISIS/WHT data we measure $FWHM_{\rm broad}$~=~$852$$\pm$$167$~kms$^{-1}$. In addition, we estimate the maximum velocity associated to the ionized gas, $V_{\rm max}$~=~|$\Delta$$V$|~+~$FWHM_{\rm broad}/2$, where $\Delta$$V$ is the velocity difference in kms$^{-1}$ between the narrow and broad kinematic components. The values of $V_{\rm max}$ are $584$$\pm$$103$~kms$^{-1}$ and $432$$\pm$$198$~kms$^{-1}$, for the MaNGA and ISIS/WHT data, respectively. The values estimated both for $FWHM_{\rm broad}$ and for $V_{\rm max}$ in the two different datasets are fully consistent with each other, taking into account the errors, confirming the extraordinary nature of the kinematics in this region. 

Based on the measured \Ha fluxes on the MaNGA data and following the same method as in \citetalias{rodriguez_del_pino_properties_2019}, the off-nuclear \rm{H~II}~region has a total associated star formation rate ($SFR$) of 0.015~M$_{\odot}$yr$^{-1}$ (including narrow and broad components). The broad component comprises a significant fraction  of the total flux in the \Ha line, $\sim30\%$ (MaNGA) and $\sim48\%$ (ISIS/WHT). 

\begin{figure}
	\begin{center}
		\includegraphics[width=0.49\textwidth]{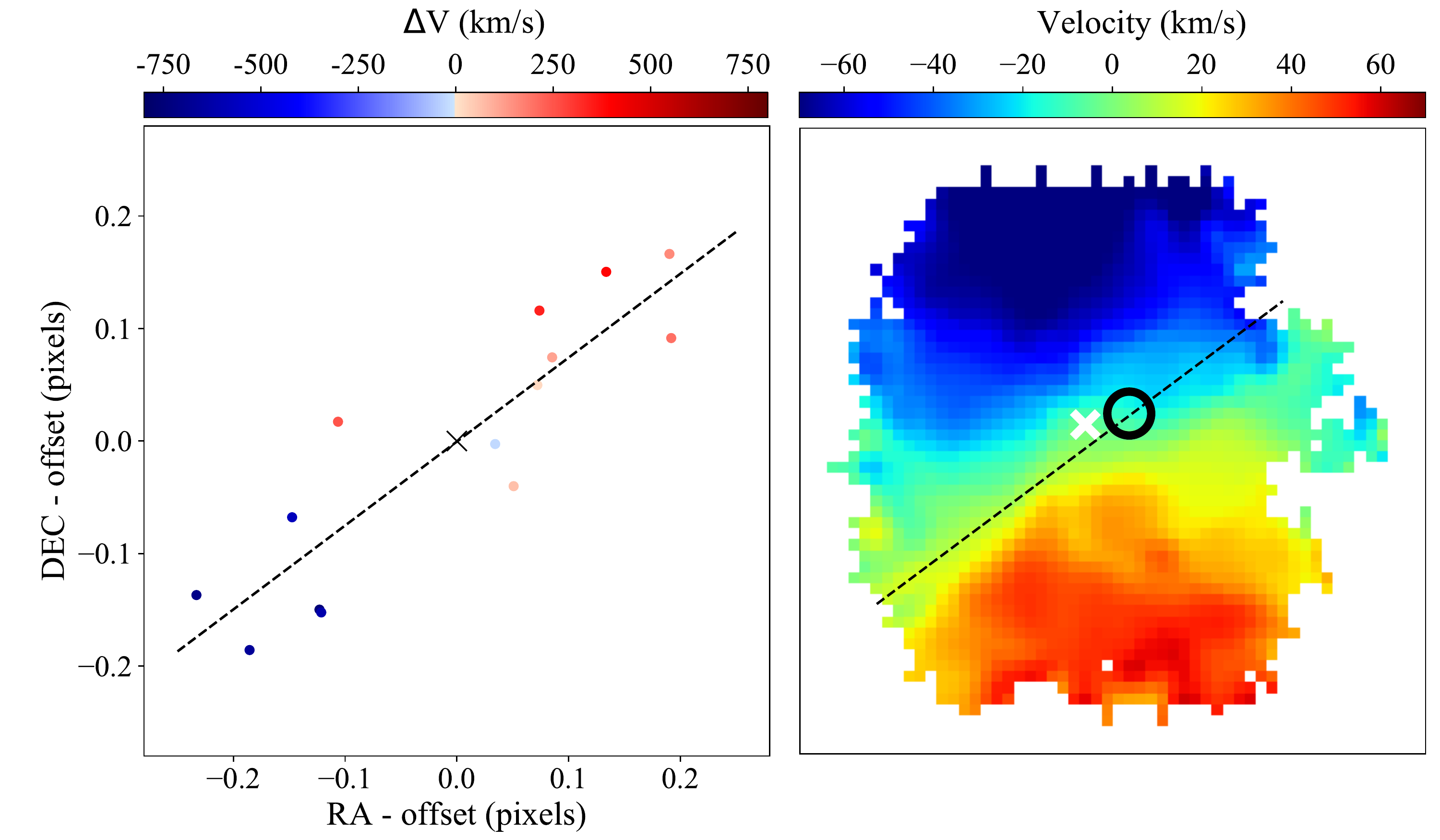}
		\caption[differential_emission]{\emph{Left}: Positions of the centroid of emission at different velocities (wavelengths) around the \rm{H~II}~region in the MaNGA data. The different spatial location of the receding (red) and preceding (blue) light indicate that the emission is spatially resolved, corresponding to a size of $\sim90$ pc. The black cross indicates the centroid of emission at the wavelength of reference, 6685\AA~($\Delta$V~=~0). \emph{Right}: Velocity field of galaxy SDSS~J143245.98+404300.3. The black circle marks the location of the \rm{H~II}~region. The dashed line corresponds to the orientation of the high velocity gas, as determined in the \emph{left} panel. As in Figure~\ref{Halphamap} the white cross marks the photometric center of the galaxy.}
		
		\label{size_estimation_plot}
	\end{center}
\vspace{-0.8cm}	
\end{figure}

\subsection{Spatial extent and orientation of the broad emission}
\label{subsection_spatial}

To measure the spatial extent of the broad emission we take into account the fact that due to the velocities of the ionized gas, the light distribution observed at different wavelengths (velocities) can be slightly shifted with respect to each other \citep{carniani_ionised_2015}. Thus, by estimating the centroid of the light distribution of the broad component at different wavelengths (velocities) we could estimate a minimum extent of the high velocity gas. For this purpose, we use the MaNGA data to generate monochromatic images of the \rm{H~II}~region at wavelengths where the broad component is dominant and that are located at both sides of the narrow \Ha emission. In our case, these spectral windows are [6668-6673]~\AA{} and [6686-6694]~\AA{} (dashed, vertical lines in Figure~\ref{spectral_fit}), and the wavelength of reference is set at 6685\AA{}. In this way, we cover a spectral range $\Delta$$\lambda$~=~26~\AA{}, which corresponds to $\sim$~1200~km/s. Then, we fit each monochromatic image with a 2D Gaussian function to find the centroid of the light distribution at each wavelength (velocity). The result of this analysis is shown in Figure~\ref{size_estimation_plot} (\emph{left} panel), where we can see that the receding light (red) is spatially shifted with respect to the light that is preceding (blue). Using the maximum distance between the points we estimate a minimum projected extent of the high velocity gas of $\sim90$~pc ($\sim0.28$~arcsec)\footnote{We note that differential atmospheric refraction could also produce a wavelength shift in the position of the centroid. For the airmass of the MaNGA observations, $\sim$1.06, taken from the MaNGA DR2 catalogue \citep{abolfathi_fourteenth_2018}, the corresponding differential atmospheric refraction between 6500\AA{} and 7000\AA{} is 0.03 arcsec \citep{filippenko_importance_1982}. Given that we are using a much smaller wavelength window, $\Delta$$\lambda$~=~26\AA{}, differential atmospheric refraction cannot produce the spatial variations of $\sim$$0.28$~arcsec that we observe.}. In the \emph{right} panel of Figure~\ref{size_estimation_plot} we show the velocity map of the host galaxy and, on top of it, the direction of the high velocity gas as obtained in the \emph{left} panel of the same figure. The orientation of the high velocity gas is slightly different to that of the projected kinematic semi-minor axis, although the kinematics in the central regions of the galaxy are quite irregular. 

\subsection{Search for broad emission in other spectral lines}
\label{subsection_search}
We explore now the presence of broad emission wings in other spectral lines available in our spectra. In the case of \NIIb, as shown in Figure~\ref{spectral_fit}, a secondary kinematic component is very weak, although its detection is complicated due to the wide wings of \Ha. Based on the results from the MCMC analysis, the probability of the broad component in \NIIb to have $S/N$~>~5 is $\sim24\%$ in MaNGA and $\sim64$\% in ISIS/WHT, and the estimated \NII/\Ha ratio for the broad component in both data sets is $\sim$0.06. The detection of broad components in $[\mathrm{S}\textsc{ii}]$$\lambda\lambda$6717,~6731 is less likely ($prob$($S/N$ > 5) < 1\%).

\begin{figure}
	\begin{center}
		\includegraphics[width=0.45\textwidth]{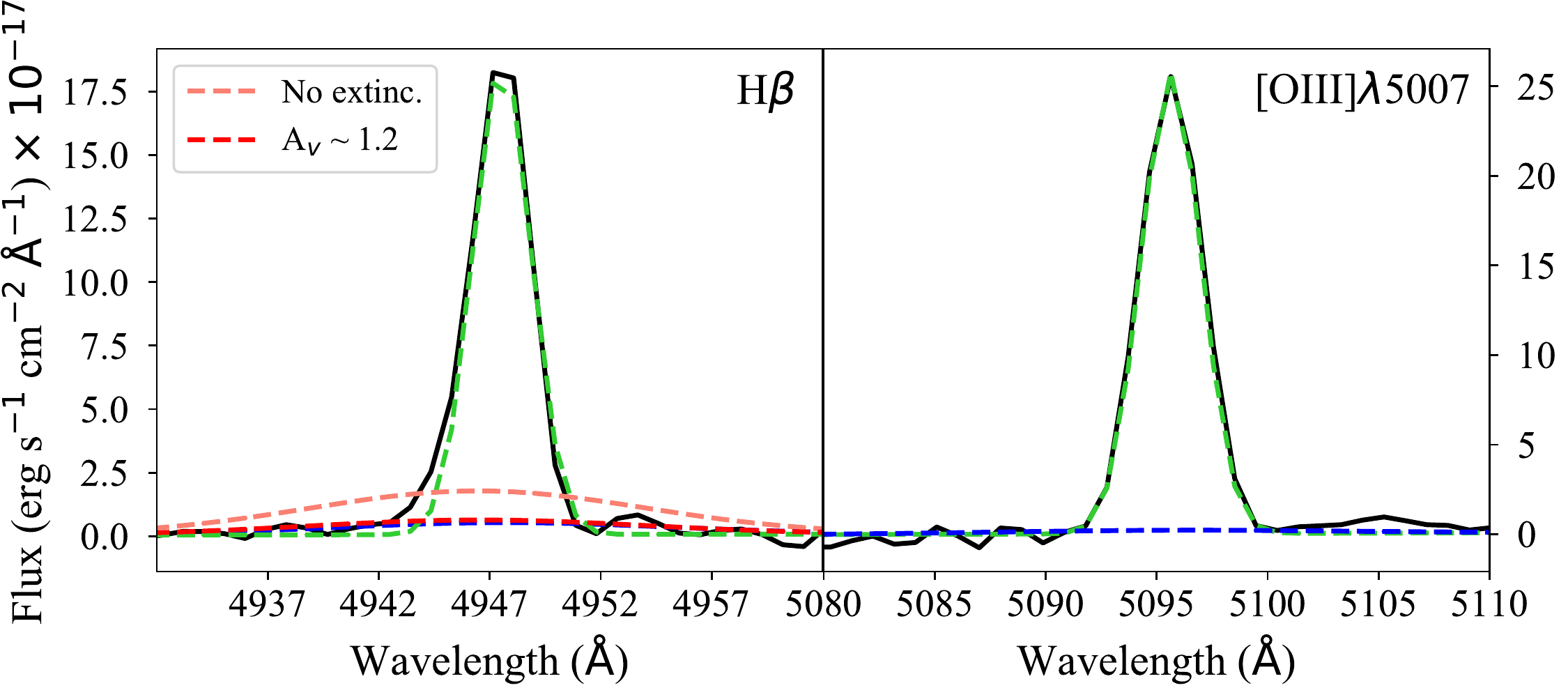}
		\caption[Halpha maps]{MaNGA spectrum of the \rm{H~II}~region around the \Hb (\emph{left}) and \OIII (\emph{right}) lines, with the result from the spectral fit (blue and green dashed lines) fixing the kinematics to those found in \Ha. In the left panel we also show the expected \Hb profile for a case with no extinction (light red) and one with $A_{\rm V}~\sim1.2$ (dark red), indicating that a moderate extinction could prevent the detection of the broad component in \Hb.
}
		\label{hbeta_oiii}
	\end{center}
\vspace*{-10pt}		
\end{figure}

For the analysis of the \Hb and \OIIIb lines we use the MaNGA spectra, once corrected for stellar absorption. To explore the presence of broad emission wings we fit these lines with a model with the same kinematics obtained in the fit to the \Ha and \NII lines, allowing the fluxes of the two kinematic components to vary. The spectral modelling of \Hb and \OIIIb is shown in Figure~\ref{hbeta_oiii}. The probabilities of detecting a broad component in \Hb and \OIIIb with $S/N$~>~5 are 39\% and 11\%, respectively, whereas the ratio \OIIIb/\Hb is $\sim0.97$. 
In Figure~\ref{hbeta_oiii} we also evaluate the effects of extinction and show that a moderate extinction, $A_{\rm V}~\sim1.2$ or larger, would be enough to erase the signature of a broad component in \Hb. In RDP19 we estimated an extinction for the narrow component of $A_{\rm V}~\sim0.3$, although the median extinction values for narrow and broad components in star-forming regions are much higher, $A_{\rm V}~\sim1.7$. 

\begin{figure*}
	\begin{center}
		\includegraphics[width=0.98\textwidth]{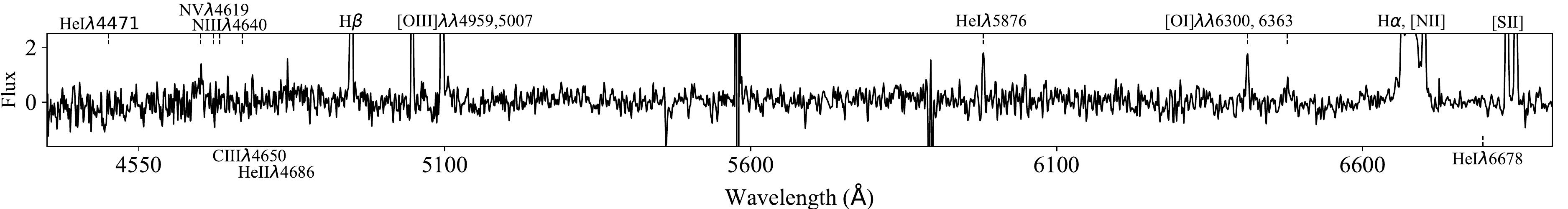}
		\caption[Halpha maps]{Full observed spectrum of the off-nuclear \rm{H~II}~region. We mark the wavelengths of the spectral lines associated to Wolf-Rayet stars (around 4700\AA{} observed wavelength) and the HeI lines, which could be indicative of broad stellar components. The lack of broad features in these spectral lines indicate that the origin of the \Ha broad wings is not related to Wolf-Rayet stars or due to contamination from supergiants or luminous blue variables.
}
		\label{plotHeIlines}
	\end{center}
\end{figure*}

In some works, the origin of broad \Ha wings in \rm{H~II}~regions has been related to the detection of a large number of Wolf-Rayet stars \citep{sargent_luminous_1991, izotov_spectrophotometry_1996}, or to contamination from broad stellar components originating in supergiants or luminous blue variables \citep[e.g.,][]{terlevich_spectroscopic_1996}. However, as shown in Figure~\ref{plotHeIlines}, the MaNGA, continuum subtracted spectrum from our off-nuclear \rm{H~II}~region does not show signatures of broad emission in spectral lines associated to Wolf-Rayet stars (\HeII, NIII $\lambda$4640 and CIII $\lambda$4650) or to the presence of supergiants or luminous blue variables (HeI lines). Following the method explained in \citet{guseva_spectroscopic_2000}, given the signal-to-noise in the spectral region around \HeII ($\sim30$) and the luminosity distance to the galaxy ($\sim80$~Mpc), we could only detect populations of more than $\sim200$ Wolf-Rayet stars, providing an upper limit on the number of this type of stars expected to be associated to our \rm{H~II}~region.

\section{Discussion and conclusions}

The extremely high velocities ($FWHM$~$\geq$~$850-1000$~kms$^{-1}$) of the ionized gas in the off-nuclear \rm{H~II}~region of SDSS J143245.98+404300.3 indicate that we are witnessing an exceptional event, specially given the low $SFR$ (0.015~M$_{\odot}$yr$^{-1}$) associated to it. In fact, such velocities are more than a factor of three larger than those observed in star-forming clumps with similar $SFRs$ in local ULIRG/s \citep{arribas_ionized_2014} and similar to those found in galaxies at $\sim2$ \citep{genzel_sins_2011}, where the $SFRs$ are much higher ($3-66$~M$_{\odot}$yr$^{-1}$). Any connection with nuclear activity is discarded given the distance ($\geq 0.8-1$~kpc) that separates the region from the center of the host galaxy. The extended nature of the region hosting gas at high velocities ($\geq90$~pc), similar to that found in other \rm{H~II}~regions \citep{castaneda_remarkable_1990, gonzalez_delgado_massive_2000}, implies that it cannot originate from a single source (e.g., supernova or AGN) and probably a large number of stellar objects are contributing. The detection of significant emission in the off-nuclear \rm{H~II}~region in the \emph{u} and \emph{g} SDSS bands (see Figure~\ref{SDSSimages}) is also consistent with the presence of a large young stellar population originated in a recent burst of star formation. The host galaxy does not show any indication of a previous interaction with nearby objects and its morphology was considered to be `uncertain' in the visual classification of Galaxy Zoo 1 \citep{lintott_galaxy_2008, lintott_galaxy_2011}. Based on the images, the galaxy is quite symmetrical, with some signs of spiral arms hosting star formation.

The fact that the high velocity gas moves, in a well-defined orientation, at positive (redshifted) and negative (blueshifted) velocities with respect to the systemic emission could be interpreted as outflowing gas that moves preferentially perpendicular to the galactic disc due to the presence of surrounding material, only allowing the gas to move in directions with less obstacles. In such case, we could evaluate the local impact of the outflow by comparing the rate at which gas is being entrained by the outflow, i.e., the outflowing mass rate, $\dot{M}$, with the rate at which gas is being converted into stars, i.e., the $SFR$ in the region. The ratio between these two parameters is the so-called "mass-loading factor", $\eta = \dot{M}/SFR$. To estimate approximated values of $\eta$ we follow the procedure explained in \citet[][equations 2-6]{arribas_ionized_2014}, assuming that the gas moves at maximum velocity and an electron density of the outflowing gas of $\sim300$~cm$^{-3}$ (median value in star-forming galaxies found in \citetalias{rodriguez_del_pino_properties_2019}). We derive $\eta$ values of $\sim9$ and $\sim7$ for the MaNGA and ISIS/WHT data, respectively, indicating that the outflow is probably producing significant quenching of star formation. The $\eta$ values derived here are extremely high, larger than those found in local star-forming galaxies \citepalias{rodriguez_del_pino_properties_2019}, local ULIRGs \citep{arribas_ionized_2014}, dwarf, starburst galaxies \citep{martin_properties_1999} and $z\sim2$ star-forming galaxies \citep{genzel_sins_2011}, which demonstrates the outstanding nature of the high velocity gas studied here.

Despite the detection of a secondary, broad kinematic component in the \Ha line, no other emission line in our spectra show clear signs of broad emission. Although the probability of detecting a broad component in \Hb, \OIIIb and \NIIb is low (see Section~\ref{subsection_search}), their associated fluxes would be consistent with ionization coming from star formation although with very low metallicities, from their position in the traditional BPT diagram \citep{baldwin_classification_1981}. A lack of broad components in $[\mathrm{N}\textsc{ii}]$$\lambda\lambda$6548,~6584 and $[\mathrm{S}\textsc{ii}]$$\lambda\lambda$6717,~6731 was also observed in NGC~2363, which was suggested to be produced by a high ionization parameter (leading to gas dominated by high excitation species) associated to the presence of turbulent mixing layers originated by the impact of fast-flowing winds \citep{binette_broad_2009}. However, there is still no consensus on what could be the origin of such strong winds. In many cases, the origin of these high velocities has been associated to the presence of a large number of Wolf-Rayet stars \citep{sargent_luminous_1991, izotov_spectrophotometry_1996, gonzalez_delgado_massive_2000}, since these stars can experience strong mass losses and large terminal velocities \citep[e.g.,][]{dodorico_wolf-rayet_1981}. However, in other cases where no Wolf-Rayet signatures are detected, a large number of massive O, B stars and/or supernovae are invoked to explain the large velocities observed \citep{castaneda_remarkable_1990, binette_broad_2009}. In our case, the lack of broad emission around the \HeII, NIII $\lambda$4640 and CIII $\lambda$4650 lines points towards a lack of a significant number of Wolf-Rayet stars, therefore indicating that the origin of the gas at very high velocities is probably related to stellar winds from a large number of massive O, B stars and/or supernovae originated in a recent burst of star formation. 

In summary, in this work we have analyzed in detail the properties of the high velocity gas ($FWHM$~$\geq$~$850-1000$~kms$^{-1}$) in an off-nuclear \rm{H~II}~region of the galaxy SDSS~J143245.98+404300.3 reported in \citetalias{rodriguez_del_pino_properties_2019}. First discovered using MaNGA data, we have further confirmed the presence of the high velocity gas using data from ISIS/WHT. The gas kinematics are much higher than those observed in local $\mathrm{H II}$\xspace~regions and similar to those in galaxies at $z\sim2$. Our analysis also indicates that the high velocity gas is probably associated to an ionized outflow entraining gas at a rate $\sim7-9$ times larger than the rate at which gas is being converted into stars in that region. Both the kinematics of the ionized gas and its off-nuclear origin make this a very singular event, with only a few similar cases being previously reported in the literature. The lack of connection to nuclear starburst or AGN activity and to Wolf-Rayet stars suggest that the origin of the large kinematics observed is connected to the presence of a large number of massive stars and/or supernovae, although the actual origin is still unknown.

\begin{acknowledgements}
We acknowledge the anonymous referee for constructive comments and suggestions. BRP, SA and AC acknowledge support from the Spanish Ministerio de
Econom\'ia y Competitividad through the grant ESP 2015-68964-P and ESP2017-83197. JPL acknowledges support from the Spanish Ministry of Economy and Competitiveness through grant AYA2017-85170-R. This work is based on observations taken with ISIS/WHT during the nights of March 9th and 10th of 2019 in visitor mode as part of the observing programme 064-WHT6/19A. Funding for the Sloan Digital Sky Survey IV has been provided by the Alfred P. Sloan Foundation, the U.S. Department of Energy Office of Science, and the Participating Institutions. SDSS-IV acknowledges support and resources from the Center for High-Performance Computing at the University of Utah. The SDSS web site is www.sdss.org. This research made use of Astropy, a community-developed core Python package for Astronomy \citep{astropy_collaboration_astropy_2018}. This research has made use of the NASA/IPAC Extragalactic Database (NED) which is operated by the Jet Propulsion Laboratory, California Institute of Technology, under contract with the National Aeronautics and Space Administration.
\end{acknowledgements}

%
%
\bibliographystyle{aa}
\bibliography{refs}

\begin{thebibliography}{29}
\expandafter\ifx\csname natexlab\endcsname\relax\def\natexlab#1{#1}\fi

\bibitem[{Abolfathi {et~al.}(2018)Abolfathi, Aguado, Aguilar, Allende~Prieto,
  Almeida, Tasnim~Ananna, Anders, Anderson, Andrews, Anguiano,
  Aragón-Salamanca, Argudo-Fernández, Armengaud, Ata, Aubourg, Avila-Reese,
  Badenes, Bailey, Balland, Barger, Barrera-Ballesteros, Bartosz, Bastien,
  Bates, Baumgarten, Bautista, Beaton, Beers, Belfiore, Bender, Bernardi,
  Bershady, Beutler, Bird, Bizyaev, Blanc, Blanton, Blomqvist, Bolton, Boquien,
  Borissova, Bovy, Andres Bradna~Diaz, Nielsen~Brandt, Brinkmann, Brownstein,
  Bundy, Burgasser, Burtin, Busca, Cañas, Cano-Díaz, Cappellari, Carrera,
  Casey, Cervantes~Sodi, Chen, Cherinka, Chiappini, Doohyun~Choi, Chojnowski,
  Chuang, Chung, Clerc, Cohen, Comerford, Comparat, Correa~do Nascimento,
  da~Costa, Cousinou, Covey, Crane, Cruz-Gonzalez, Cunha, da~Silva~Ilha, Damke,
  Darling, Davidson, Dawson, de~Icaza~Lizaola, de~la Macorra, de~la Torre,
  De~Lee, de~Sainte~Agathe, Deconto~Machado, Dell’Agli, Delubac,
  Diamond-Stanic, Donor, José~Downes, Drory, du~Mas~des Bourboux, Duckworth,
  Dwelly, Dyer, Ebelke, Davis~Eigenbrot, Eisenstein, Elsworth, Emsellem,
  Eracleous, Erfanianfar, Escoffier, Fan, Fernández~Alvar, Fernandez-Trincado,
  Cirolini, Feuillet, Finoguenov, Fleming, Font-Ribera, Freischlad, Frinchaboy,
  Fu, Gómez Maqueo~Chew, Galbany, García~Pérez, Garcia-Dias,
  García-Hernández, Garma~Oehmichen, Gaulme, Gelfand, Gil-Marín, Gillespie,
  Goddard, González~Hernández, Gonzalez-Perez, Grabowski, Green, Grier,
  Gueguen, Guo, Guy, Hagen, Hall, Harding, Hasselquist, Hawley, Hayes, Hearty,
  Hekker, Hernandez, Hernandez~Toledo, Hogg, Holley-Bockelmann, Holtzman, Hou,
  Hsieh, Hunt, Hutchinson, Hwang, Jimenez~Angel, Johnson, Jones, Jönsson,
  Jullo, Sakil~Khan, Kinemuchi, Kirkby, Kirkpatrick, Kitaura, Knapp, Kneib,
  Kollmeier, Lacerna, Lane, Lang, Law, Le~Goff, Lee, Li, Li, Lian, Liang, Lima,
  Lin, Long, Lucatello, Lundgren, Mackereth, MacLeod, Mahadevan, Geimba~Maia,
  Majewski, Manchado, Maraston, Mariappan, Marques-Chaves, Masseron, Masters,
  McDermid, McGreer, Melendez, Meneses-Goytia, Merloni, Merrifield, Meszaros,
  Meza, Minchev, Minniti, Mueller, Muller-Sanchez, Muna, Muñoz, Myers, Nair,
  Nandra, Ness, Newman, Nichol, Nidever, Nitschelm, Noterdaeme, O’Connell,
  Oelkers, Oravetz, Oravetz, Aquino~Ortíz, Osorio, Pace, Padilla,
  Palanque-Delabrouille, Alonso~Palicio, Pan, Pan, Parikh, Pâris, Park,
  Peirani, Pellejero-Ibanez, Penny, Percival, Perez-Fournon, Petitjean, Pieri,
  Pinsonneault, Pisani, Prada, Prakash, Queiroz, Raddick, Raichoor,
  Barboza~Rembold, Richstein, Riffel, Riffel, Rix, Robin, Rodríguez~Torres,
  Román-Zúñiga, Ross, Rossi, Ruan, Ruggeri, Ruiz, Salvato, Sánchez,
  Sánchez, Sanchez~Almeida, Sánchez-Gallego, Santana~Rojas, Santiago,
  Schiavon, Schimoia, Schlafly, Schlegel, Schneider, Schuster, Schwope, Seo,
  Serenelli, Shen, Shen, Shetrone, Shull, Silva~Aguirre, Simon, Skrutskie,
  Slosar, Smethurst, Smith, Sobeck, Somers, Souter, Souto, Spindler, Stark,
  Stassun, Steinmetz, Stello, Storchi-Bergmann, Streblyanska, Stringfellow,
  Suárez, Sun, Szigeti, Taghizadeh-Popp, Talbot, Tang, Tao, Tayar, Tembe,
  Teske, Thakar, Thomas, Tissera, Tojeiro, Tremonti, Troup, Urry, Valenzuela,
  van~den Bosch, Vargas-González, Vargas-Magaña, Vazquez, Villanova, Vogt,
  Wake, Wang, Weaver, Weijmans, Weinberg, Westfall, Whelan, Wilcots, Wild,
  Williams, Wilson, Wood-Vasey, Wylezalek, Xiao, Yan, Yang, Ybarra, Yèche,
  Zakamska, Zamora, Zarrouk, Zasowski, Zhang, Zhao, Zhao, Zheng, Zheng, Zhou,
  Zhu, Zinn, \& Zou}]{abolfathi_fourteenth_2018}
Abolfathi, B., Aguado, D.~S., Aguilar, G., {et~al.} 2018, The Astrophysical
  Journal Supplement Series, 235, 42

\bibitem[{Amorín {et~al.}(2012)Amorín, Vílchez, Hägele, Firpo,
  Pérez-Montero, \& Papaderos}]{amorin_complex_2012}
Amorín, R., Vílchez, J.~M., Hägele, G.~F., {et~al.} 2012, The Astrophysical
  Journal Letters, 754, L22

\bibitem[{Arribas {et~al.}(2014)Arribas, Colina, Bellocchi, Maiolino, \&
  Villar-Martín}]{arribas_ionized_2014}
Arribas, S., Colina, L., Bellocchi, E., Maiolino, R., \& Villar-Martín, M.
  2014, Astronomy and Astrophysics, 568, A14

\bibitem[{{Astropy Collaboration} {et~al.}(2018){Astropy Collaboration},
  Price-Whelan, Sipőcz, Günther, Lim, Crawford, Conseil, Shupe, Craig,
  Dencheva, Ginsburg, VanderPlas, Bradley, Pérez-Suárez, de~Val-Borro,
  Aldcroft, Cruz, Robitaille, Tollerud, Ardelean, Babej, Bach, Bachetti,
  Bakanov, Bamford, Barentsen, Barmby, Baumbach, Berry, Biscani, Boquien,
  Bostroem, Bouma, Brammer, Bray, Breytenbach, Buddelmeijer, Burke, Calderone,
  Cano~Rodríguez, Cara, Cardoso, Cheedella, Copin, Corrales, Crichton,
  D'Avella, Deil, Depagne, Dietrich, Donath, Droettboom, Earl, Erben, Fabbro,
  Ferreira, Finethy, Fox, Garrison, Gibbons, Goldstein, Gommers, Greco,
  Greenfield, Groener, Grollier, Hagen, Hirst, Homeier, Horton, Hosseinzadeh,
  Hu, Hunkeler, Ivezić, Jain, Jenness, Kanarek, Kendrew, Kern, Kerzendorf,
  Khvalko, King, Kirkby, Kulkarni, Kumar, Lee, Lenz, Littlefair, Ma, Macleod,
  Mastropietro, McCully, Montagnac, Morris, Mueller, Mumford, Muna, Murphy,
  Nelson, Nguyen, Ninan, Nöthe, Ogaz, Oh, Parejko, Parley, Pascual, Patil,
  Patil, Plunkett, Prochaska, Rastogi, Reddy~Janga, Sabater, Sakurikar,
  Seifert, Sherbert, Sherwood-Taylor, Shih, Sick, Silbiger, Singanamalla,
  Singer, Sladen, Sooley, Sornarajah, Streicher, Teuben, Thomas, Tremblay,
  Turner, Terrón, van Kerkwijk, de~la Vega, Watkins, Weaver, Whitmore,
  Woillez, Zabalza, \& {Astropy
  Contributors}}]{astropy_collaboration_astropy_2018}
{Astropy Collaboration}, Price-Whelan, A.~M., Sipőcz, B.~M., {et~al.} 2018,
  The Astronomical Journal, 156, 123

\bibitem[{Baldwin {et~al.}(1981)Baldwin, Phillips, \&
  Terlevich}]{baldwin_classification_1981}
Baldwin, J.~A., Phillips, M.~M., \& Terlevich, R. 1981, Publications of the
  Astronomical Society of the Pacific, 93, 5

\bibitem[{Binette {et~al.}(2009)Binette, Drissen, Ubeda, Raga, Robert, \&
  Krongold}]{binette_broad_2009}
Binette, L., Drissen, L., Ubeda, L., {et~al.} 2009, Astronomy and Astrophysics,
  500, 817

\bibitem[{Bundy {et~al.}(2015)Bundy, Bershady, Law, Yan, Drory, MacDonald,
  Wake, Cherinka, Sánchez-Gallego, Weijmans, Thomas, Tremonti, Masters,
  Coccato, Diamond-Stanic, Aragón-Salamanca, Avila-Reese, Badenes,
  Falcón-Barroso, Belfiore, Bizyaev, Blanc, Bland-Hawthorn, Blanton,
  Brownstein, Byler, Cappellari, Conroy, Dutton, Emsellem, Etherington,
  Frinchaboy, Fu, Gunn, Harding, Johnston, Kauffmann, Kinemuchi, Klaene,
  Knapen, Leauthaud, Li, Lin, Maiolino, Malanushenko, Malanushenko, Mao,
  Maraston, McDermid, Merrifield, Nichol, Oravetz, Pan, Parejko, Sanchez,
  Schlegel, Simmons, Steele, Steinmetz, Thanjavur, Thompson, Tinker, van~den
  Bosch, Westfall, Wilkinson, Wright, Xiao, \& Zhang}]{bundy_overview_2015}
Bundy, K., Bershady, M.~A., Law, D.~R., {et~al.} 2015, The Astrophysical
  Journal, 798, 7

\bibitem[{Carniani {et~al.}(2015)Carniani, Marconi, Maiolino, Balmaverde,
  Brusa, Cano-Díaz, Cicone, Comastri, Cresci, Fiore, Feruglio, Franca,
  Mainieri, Mannucci, Nagao, Netzer, Piconcelli, Risaliti, Schneider, \&
  Shemmer}]{carniani_ionised_2015}
Carniani, S., Marconi, A., Maiolino, R., {et~al.} 2015, Astronomy \&
  Astrophysics, 580, A102

\bibitem[{Casta\~{n}eda {et~al.}(1990)Casta\~{n}eda, Vilchez, \&
  Copetti}]{castaneda_remarkable_1990}
Casta\~{n}eda, H.~O., Vilchez, J.~M., \& Copetti, M. V.~F. 1990, The
  Astrophysical Journal, 365, 164

\bibitem[{Diaz {et~al.}(1987)Diaz, Terlevich, Pagel, Vilchez, \&
  Edmunds}]{diaz_detailed_1987}
Diaz, A.~I., Terlevich, E., Pagel, B. E.~J., Vilchez, J.~M., \& Edmunds, M.~G.
  1987, Monthly Notices of the Royal Astronomical Society, 226, 19

\bibitem[{Dodorico \& Rosa(1981)}]{dodorico_wolf-rayet_1981}
Dodorico, S. \& Rosa, M. 1981, The Astrophysical Journal, 248, 1015

\bibitem[{Filippenko(1982)}]{filippenko_importance_1982}
Filippenko, A.~V. 1982, Publications of the Astronomical Society of the
  Pacific, 94, 715

\bibitem[{Foreman-Mackey {et~al.}(2013)Foreman-Mackey, Hogg, Lang, \&
  Goodman}]{foreman-mackey_emcee:_2013}
Foreman-Mackey, D., Hogg, D.~W., Lang, D., \& Goodman, J. 2013, Publications of
  the Astronomical Society of the Pacific, 125, 306, arXiv: 1202.3665

\bibitem[{Genzel {et~al.}(2011)Genzel, Newman, Jones, Förster~Schreiber,
  Shapiro, Genel, Lilly, Renzini, Tacconi, Bouché, Burkert, Cresci, Buschkamp,
  Carollo, Ceverino, Davies, Dekel, Eisenhauer, Hicks, Kurk, Lutz, Mancini,
  Naab, Peng, Sternberg, Vergani, \& Zamorani}]{genzel_sins_2011}
Genzel, R., Newman, S., Jones, T., {et~al.} 2011, The Astrophysical Journal,
  733, 101

\bibitem[{González~Delgado \& Pérez(2000)}]{gonzalez_delgado_massive_2000}
González~Delgado, R.~M. \& Pérez, E. 2000, Monthly Notices of the Royal
  Astronomical Society, 317, 64

\bibitem[{Guseva {et~al.}(2000)Guseva, Izotov, \&
  Thuan}]{guseva_spectroscopic_2000}
Guseva, N.~G., Izotov, Y.~I., \& Thuan, T.~X. 2000, The Astrophysical Journal,
  531, 776

\bibitem[{Heckman \& Borthakur(2016)}]{heckman_implications_2016}
Heckman, T.~M. \& Borthakur, S. 2016, The Astrophysical Journal, 822, 9

\bibitem[{Izotov {et~al.}(1996)Izotov, Dyak, Chaffee, Foltz, Kniazev, \&
  Lipovetsky}]{izotov_spectrophotometry_1996}
Izotov, Y.~I., Dyak, A.~B., Chaffee, F.~H., {et~al.} 1996, The Astrophysical
  Journal, 458, 524

\bibitem[{Lintott {et~al.}(2011)Lintott, Schawinski, Bamford, Slosar, Land,
  Thomas, Edmondson, Masters, Nichol, Raddick, Szalay, Andreescu, Murray, \&
  Vandenberg}]{lintott_galaxy_2011}
Lintott, C., Schawinski, K., Bamford, S., {et~al.} 2011, Monthly Notices of the
  Royal Astronomical Society, 410, 166

\bibitem[{Lintott {et~al.}(2008)Lintott, Schawinski, Slosar, Land, Bamford,
  Thomas, Raddick, Nichol, Szalay, Andreescu, Murray, \&
  Vandenberg}]{lintott_galaxy_2008}
Lintott, C.~J., Schawinski, K., Slosar, A., {et~al.} 2008, Monthly Notices of
  the Royal Astronomical Society, 389, 1179

\bibitem[{Martin(1999)}]{martin_properties_1999}
Martin, C.~L. 1999, The Astrophysical Journal, 513, 156

\bibitem[{{Planck Collaboration} {et~al.}(2014){Planck Collaboration}, Ade,
  Aghanim, Armitage-Caplan, Arnaud, Ashdown, Atrio-Barandela, Aumont,
  Baccigalupi, Banday, Barreiro, Bartlett, Battaner, Benabed, Benoît,
  Benoit-Lévy, Bernard, Bersanelli, Bielewicz, Bobin, Bock, Bonaldi, Bond,
  Borrill, Bouchet, Bridges, Bucher, Burigana, Butler, Calabrese, Cappellini,
  Cardoso, Catalano, Challinor, Chamballu, Chary, Chen, Chiang, Chiang,
  Christensen, Church, Clements, Colombi, Colombo, Couchot, Coulais, Crill,
  Curto, Cuttaia, Danese, Davies, Davis, de~Bernardis, de~Rosa, de~Zotti,
  Delabrouille, Delouis, Désert, Dickinson, Diego, Dolag, Dole, Donzelli,
  Doré, Douspis, Dunkley, Dupac, Efstathiou, Elsner, Enßlin, Eriksen,
  Finelli, Forni, Frailis, Fraisse, Franceschi, Gaier, Galeotta, Galli, Ganga,
  Giard, Giardino, Giraud-Héraud, Gjerløw, González-Nuevo, Górski, Gratton,
  Gregorio, Gruppuso, Gudmundsson, Haissinski, Hamann, Hansen, Hanson,
  Harrison, Henrot-Versillé, Hernández-Monteagudo, Herranz, Hildebrandt,
  Hivon, Hobson, Holmes, Hornstrup, Hou, Hovest, Huffenberger, Jaffe, Jaffe,
  Jewell, Jones, Juvela, Keihänen, Keskitalo, Kisner, Kneissl, Knoche, Knox,
  Kunz, Kurki-Suonio, Lagache, Lähteenmäki, Lamarre, Lasenby, Lattanzi,
  Laureijs, Lawrence, Leach, Leahy, Leonardi, León-Tavares, Lesgourgues,
  Lewis, Liguori, Lilje, Linden-Vørnle, López-Caniego, Lubin, Macías-Pérez,
  Maffei, Maino, Mandolesi, Maris, Marshall, Martin, Martínez-González, Masi,
  Massardi, Matarrese, Matthai, Mazzotta, Meinhold, Melchiorri, Melin, Mendes,
  Menegoni, Mennella, Migliaccio, Millea, Mitra, Miville-Deschênes, Moneti,
  Montier, Morgante, Mortlock, Moss, Munshi, Murphy, Naselsky, Nati, Natoli,
  Netterfield, Nørgaard-Nielsen, Noviello, Novikov, Novikov, O'Dwyer, Osborne,
  Oxborrow, Paci, Pagano, Pajot, Paladini, Paoletti, Partridge, Pasian,
  Patanchon, Pearson, Pearson, Peiris, Perdereau, Perotto, Perrotta, Pettorino,
  Piacentini, Piat, Pierpaoli, Pietrobon, Plaszczynski, Platania,
  Pointecouteau, Polenta, Ponthieu, Popa, Poutanen, Pratt, Prézeau, Prunet,
  Puget, Rachen, Reach, Rebolo, Reinecke, Remazeilles, Renault, Ricciardi,
  Riller, Ristorcelli, Rocha, Rosset, Roudier, Rowan-Robinson, Rubiño-Martín,
  Rusholme, Sandri, Santos, Savelainen, Savini, Scott, Seiffert, Shellard,
  Spencer, Starck, Stolyarov, Stompor, Sudiwala, Sunyaev, Sureau, Sutton,
  Suur-Uski, Sygnet, Tauber, Tavagnacco, Terenzi, Toffolatti, Tomasi, Tristram,
  Tucci, Tuovinen, Türler, Umana, Valenziano, Valiviita, Van~Tent, Vielva,
  Villa, Vittorio, Wade, Wandelt, Wehus, White, White, Wilkinson, Yvon,
  Zacchei, \& Zonca}]{planck_collaboration_planck_2014}
{Planck Collaboration}, Ade, P. A.~R., Aghanim, N., {et~al.} 2014, Astronomy
  and Astrophysics, 571, A16

\bibitem[{Rodríguez~del Pino {et~al.}(2019)Rodríguez~del Pino, Arribas,
  Piqueras~López, Villar-Martín, \&
  Colina}]{rodriguez_del_pino_properties_2019}
Rodríguez~del Pino, B., Arribas, S., Piqueras~López, J., Villar-Martín, M.,
  \& Colina, L. 2019, Monthly Notices of the Royal Astronomical Society, 486,
  344

\bibitem[{Roy {et~al.}(1992)Roy, Aube, McCall, \& Dufour}]{roy_origin_1992}
Roy, J.-R., Aube, M., McCall, M.~L., \& Dufour, R.~J. 1992, The Astrophysical
  Journal, 386, 498

\bibitem[{Roy {et~al.}(1991)Roy, Boulesteix, Joncas, \&
  Grundseth}]{roy_superbubble_1991}
Roy, J.-R., Boulesteix, J., Joncas, G., \& Grundseth, B. 1991, The
  Astrophysical Journal, 367, 141

\bibitem[{Sargent \& Filippenko(1991)}]{sargent_luminous_1991}
Sargent, W. L.~W. \& Filippenko, A.~V. 1991, The Astronomical Journal, 102, 107

\bibitem[{Terlevich {et~al.}(1996)Terlevich, Díaz, Terlevich,
  González-Delgado, Pérez, \& García~Vargas}]{terlevich_spectroscopic_1996}
Terlevich, E., Díaz, A.~I., Terlevich, R., {et~al.} 1996, Monthly Notices of
  the Royal Astronomical Society, 279, 1219

\bibitem[{Terlevich {et~al.}(2014)Terlevich, Terlevich, Bosch, Díaz, Hägele,
  Cardaci, \& Firpo}]{terlevich_high-velocity_2014}
Terlevich, R., Terlevich, E., Bosch, G., {et~al.} 2014, Monthly Notices of the
  Royal Astronomical Society, 445, 1449

\bibitem[{Villar~Martín {et~al.}(2014)Villar~Martín, Emonts, Humphrey,
  Cabrera~Lavers, \& Binette}]{villar_martin_triggering_2014}
Villar~Martín, M., Emonts, B., Humphrey, A., Cabrera~Lavers, A., \& Binette,
  L. 2014, Monthly Notices of the Royal Astronomical Society, 440, 3202

\end{thebibliography}

\end{document}